\documentstyle[epsf,aps]{revtex}

\twocolumn

\begin{document}

\title{ Analogy between optimal spin estimation and
interferometry }

\author{Zden\v{e}k Hradil and Miloslav Du\v{s}ek}

\address{ Department of Optics, Palack\'{y}
University, 17. listopadu 50, 772 00 Olomouc, Czech Republic}

\date{DRAFT:~\today}

\maketitle

\begin{abstract}

Scheme for optimal  spin state estimation is considered in analogy with
phase detection in interferometry. Recently reported coherent
measurements yielding the average fidelity $(N+1)/(N+2)$ for $N$
particle system corresponds to the standard limit of phase resolution
$1/\sqrt{N}. $ It provides the bound  for incoherent measurements when
each particle is detected separately and information is used optimally.
For specific states, improvement up to the value $1/N$ is possible in
quantum theory. The best results are obtained combining sequentially
coherent measurements on fractional groups of particles.

\end{abstract}

\draft

\section{Introduction}

Recently  Peres and Wotters \cite{PW} formulated a
conjecture: Coherent measurement performed on the collective system
is more efficient than sequential measurements of individual particles.
This idea has been further developed
by Massar and Popescu \cite{MassarP}.  They formulated this
conjecture as a proposal for  a ``quantum game."  The player has
$N$ identical copies of    1/2   spin particles prepared in an unknown pure
state  $|\psi\rangle $, and he is allowed to do any measurement.
Possible results of the
measurement will be denoted by an index $r.$
The aim of the measurement is to determine the original   state
of the system. Therefore,  in the next step the measured data should
be attached to a state $|\psi\rangle_r$, which represents the players's
estimation of an  unknown
true  state.
 In the last stage of the  game the true state is compared with
its estimate and their coincidence is quantified
by the so-called fidelity:
$|\langle \psi|\psi\rangle_r|^2.$ The runs are  repeated many times
with varying  true state $|\psi\rangle.$  The final score
of the quantum game is given by the averaged  fidelity
 \begin{equation}
 S = \{  |\langle \psi|\psi\rangle_r|^2 \}_{ av },
 \end{equation}
where averaging is carried out over the measured data and all
the true states $\{\}_{av} =
\{r,|\psi \rangle\}. $
 Massar and Popescu proved  that the
maximum score is   $(N+1)/(N+2).$
This value cannot  be reached
by measurements acting on isolated particles.
Derka, Bu\v{z}ek and
Ekert \cite{Buzek} showed that this score can be obtained by coherent
measurement described by a finite-dimensional probability operator
valued measure.

The aim of this contribution is to  address the relation  between
recently optimized measurement, repeated measurement
on the   Stern-Gerlach apparatus and interferometry.
 Particularly, it will be demonstrated that the above
mentioned  score represents the  ultimate limitation  for
sequential measurements performed on each particle separately
for any quantum state.
This  corresponds to the  standard resolution  $1/\sqrt{N}$
currently reached in  interferometry.
However, this regime does not represent  the ultimate strategy.
In analogy with quantum interferometry performance may achieve
the resolution up to   $1/N$ provided
that spin orientation is properly coded into a quantum state
of $N$ particles.
In the particular
case addressed in this contribution the optimal strategy corresponds  to
sequentially performed  coherent  measurements. This explicit example
demonstrates the complexity of optimal measurement, which can combine
advantages of both the coherent and sequential measurements with groups of
particles.

\section{Adaptive Stern-Gerlach  spin detection }

Assume a standard measurement on an ideal Stern-Gerlach apparatus.
A sample  particle is  prepared in the pure spin  state
\begin{equation}
 |{\bf n}\rangle \langle {\bf n}| = \frac{1}{2}[1 + ({\bf n} \sigma)],
\end{equation}
where ${\bf n}$  represents the unity vector on the Poincare
sphere, $({\bf n} \sigma ) $ being its  scalar product with
vector of  Pauli
matrices.
The   impinging particles  will be  deviated  up or down.
 In the long run of
repetitions the relative frequencies  will
approach the prediction of quantum theory.
Representing  the setting of the Stern-Gerlach apparatus
by the unity vector ${\bf m}$, the
probabilities of detection ``up'' ($+$) and ``down'' ($-$) read
\begin{equation}
 p_{\pm} = \frac{1}{2}[1 \pm ({\bf m n} )].
\end{equation}
What is the best possible but still feasible result,
which would  predict the spin
orientation  with the highest accuracy?
The most accurate state estimation may  be done
if all tested
particles were  registered in the same  output channel of the SG
apparatus.
In such a case  the best estimation of the
spin corresponds to the orientation of the SG apparatus.
Of course, it does not mean that  the estimated direction will
fit the spin orientation exactly. Deviations are distributed
according to the  posterior distribution conditioned by the
detected data.  This can be handled analogously to the phase
estimation  \cite{phase}.  The
deviations  between the estimated and the true directions are
given by  the  Bayes theorem  as  the posterior probability  density
\begin{equation}
P({\bf n})  = \frac{N+1}{4\pi} \cos^{2N}(\theta/2)
\end{equation}
over the Poincar\'e sphere.
The vector ${\bf n}$ is parametrized  by the Euler  angles
$\theta, \phi$ in the coordinate system where the direction
of $z$-axis coincide always  with the estimated  direction (i.e. with the
orientation of SG apparatus ${\bf m}$).
Notice that this is in accordance with the rules of
the quantum game as defined  in \cite{MassarP}. The result of the
measurement is always a specific direction, namely the setting of
SG apparatus. In this case the
score reads
\begin{equation}
S = \int P({\bf n}) d^2 \Omega_{{\bf n}} \cos^2(\theta/2)     =
\frac{N+1}{N+2}.
\end{equation}
This is the  upper bound  of sequential  measurements on
single  particles with felicitously rotated  SG apparatus.
This is obviously an ultimate limitation since such results  are
possible and none  measurement performed sequentially on single particles can
 yield better spin prediction.
However, on the contrary to the coherent measurements,
 this resolution cannot be  really achieved, but  it may be approximated with
an  arbitrary accuracy  for $N$ large enough.

\begin{figure}
     \epsfxsize=0.8\hsize \epsffile{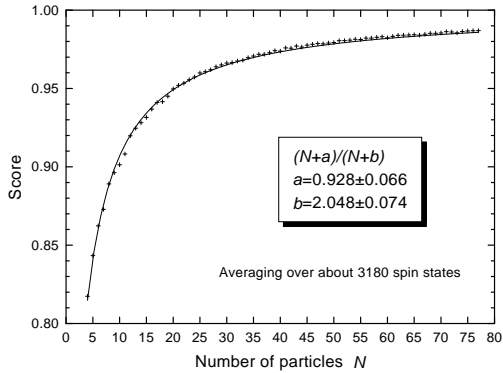} \caption{Numerical
   simulation of one-particle adaptive measurement.} \label{fig1}
\end{figure}

The possible
 realization may be suggested as an adaptive scheme, where the
orientation of SG device depends on the previous results. The aim
of the  scheme is to find such  an orientation, where almost  all the
particles are  counted on the same port. This is obviously always
a little worse than ideal case, since  some portion of
counted particles must always  be used
for corrections of the  SG-device orientation and are therefore ``lost''.
Differences between ideal  and realistic scheme seem to be
 negligible as demonstrated in Fig.~\ref{fig1}.
An adaptive measurement is simulated here. The procedure starts
with projecting a single  particle into the three orthogonal
randomly chosen  directions.
The choice of the orientation of the  subsequent SG  measurement
represents its own  interesting    optimization problem.
Obviously, for getting the best score, it seems to be
advantageous to project the particles  into the most probable
orientation. However, this will not reveal  new  corrections to
the orientation of SG apparatus well. On the other hand, the
particles may be counted with the SG device  oriented
perpendicularly to the most probable orientation. In this case the spreading
of the data will be obviously broader than in the former case,
but such measurement will be more sensitive to the deviation from the true
direction. In the simulation scheme, the latter approach has been
used and  algorithm for
synthesis of incompatible spin projections   has been used \cite{spin}.
The treatment is still not optimal, however, as will be seen
in the next paragraph, the differences
seem to be unimportant.

\section{Analogy between spin estimation and interferometry}

There is a  clear analogy between spin measurement  and
 phase  interferometry.
As the proper resolution measure, the dispersion  of phase
 may  be  defined as \cite{rao,phase2}
\begin{equation}
D^2 = 1 -  \{ \cos\theta \}_{av}^2.
\end{equation}
Usually, the averaging is done over the data only.
Provided that the width of  phase distribution is small,
dispersion tends to the  standard variance of phase.
Using  the definition of the  score
\begin{equation}
S = \frac{1}{2} + \frac{1}{2} \{ \cos\theta \}_{av},
\end{equation}
both the measures fulfil the
relation
\begin{equation}
S(S+1) = \frac{D^2}{4}.
      %S \approx 1 -\frac{D^2}{4}.
\end{equation}
Consequently, the
value $ S= (N+1)/(N+2)$  is nothing else as dispersion (phase
variance)
$ D \approx  2/\sqrt{N}.$  This is the so-called
 standard limit of phase resolution
\cite{Yurke,Milburn}. Any standard measurement is scaled in this
way and all the measurements differ by some
multiplicative factors only.  Loosely speaking, all the
classical
strategies are essentially equivalent from this viewpoint.
This is why it has perhaps little sense to optimize further the
adaptive scheme. For example,
provided that one will use the most straightforward method of
spin estimation based on the measurement of $x,y,$ and $z$ components  of the
spin on the Poincar\' e sphere always with $N/3 $ particles,
the resulting score  may be evaluated  asymptotically  as $S
\approx  1 - 114/(100 N).$
The difference  between optimal coherent and
realistic sequential measurements is assumed sometimes to be
significant for small number of particles.
However, in this case all the predictions are rather
uncertain.
In the case of phase detection
it has little sense to compare two phase distributions,
whose widths are comparable to $2\pi$  window.
Then the phase knowledge is almost equally bad.
Conventional resolution measures possess good meaning only when the
effective width is substantially  less than the width of the
interval.

%%%%%%%%
%All these arguments support our opinion that the score
%$S = (N+1)/(N+2) $ is the ultimate limit for
%classical strategies consisting of  sequential
%measurements of individual particles.
%%%%%%%%

More profound  analogy  between the spin measurement
and interferometry  follows from   the common nature of the  SU(2)
symmetry \cite{Yurke}.  As well known, the resolution up to the
order   $1/N$ may be achieved in interferometry.
In this case, however, it is recommended
to modify slightly  the proposed quantum game.
Suppose that somebody wants
to communicate  an orientation of the axis in 3D space. Just an axis, not
the direction of an ``arrow''.  For this  purpose  $N$ 1/2-spin particles
are available and any measurement on these particles is allowed.
Instead of $N$ identical copies, one may consider a
general quantum state spanned by $N$
particles, in which the unknown orientation is coded.
As the result of the measurement the unknown  axis  should be find.
The score is defined in the same manner as
before. The question are: ``How to encode information on axis
orientation into a quantum state of the particles? What measurement
should be done in order to obtain the best score?''

\section{Superresolution in interferometry and spin estimation}

Let us review briefly the  description of SU(2) interferometers.
 The  transformation of an
internal state is given  by an  unitary transformation
of an input state
\begin{eqnarray}
 \hat U(\theta, \varphi)  = \exp\left[-i \theta
 \hat{J}_2(\varphi) \right]\equiv \nonumber \\
 \exp(i \varphi \hat J_3) \exp(-i \theta \hat J_2 )\exp(-i \varphi \hat
 J_3),
\end{eqnarray}
where $\hat J_1, \hat J_2, \hat J_3$ corresponds to generators of
SU(2) group \cite{Yurke,Milburn}.
The transformation is given by  the unity vector
$ {\bf n} = (\cos\varphi
\sin\theta, \sin\varphi \sin\theta, \cos \theta).$
An input state $|{\rm in}\rangle$ may be
any  $N$ particle state.
The measurement  may be represented by projectors
into an output state   $|{\rm out}\rangle.$
The posterior phase distribution corresponds to the
 scattering amplitude
\begin{equation}
P(\theta,\varphi) \approx  |\langle {\rm in} | \hat
U(\theta,\varphi)| {\rm out}\rangle |^2.
\end{equation}

This  scheme  encompasses the  Massar, Popescu
quantum game as a special case
for the choice   of the  input state
$
 |{\rm in}\rangle = |N,N\rangle .
$
Score depends on the accuracy of  detection  of   $
\theta$  variable. As well known, the highest accuracy
is achieved when  the phase shift near the zero value is detected.
This corresponds to the detection of  the same quantum state on the output
$
 |{\rm out}\rangle = |N,N\rangle .
$
 The particles  feeding a
single input port of an interferometer    appears again   on the
corresponding  output port. Obviously, the interpretation does
not change  provided that  particles will enter and leave the
interferometer sequentially. This is the consequence of the
famous Dirac statement that  ``each particle
interferes with itself."  \cite{Dirac}
Ultimate score $(N+1)/(N+2)$ is relevant just to this regime.
Considerations related to adaptive SG detection are obviously
related also to this situation.

In the following, for given input state and for given setting
of the parameters of the  interferometer, only the most favourable but
still feasible output will be assumed. Such felicitous results provide an
upper bound  conditioned by the realistic  measurements. Obviously, no
other measurement of the given type, can provide better prediction.

As well known in interferometry \cite{Yurke,Milburn,Holland},
better resolution may be
obtained  provided that both the input ports of an interferometer
are fed simultaneously by an equal number of $r$  particles.
In the  case when the same output state appears on the output,
the phase shift prediction will be the sharpest.
This corresponds to the input and output states
$
 |{\rm in}\rangle = |2r,0\rangle
$
and
$
 |{\rm out}\rangle = |2r,0\rangle .
$
The scattering amplitude then depends only on the variable
$\theta$ as
\begin{equation}
P(\theta) \propto \left[P^0_r(\cos\theta)\right]^2.
\label{Leg}
\end{equation}
The Legendre  function $P^0_r(\cos\theta)$ may be approximated by Bessel
function $J_0(r\theta)$ for large index $r$. Consequently, the
probability distribution is not integrable for $r \rightarrow \infty$
and must be therefore treated more carefully \cite{acta}. Particularly,
it is therefore not the best strategy to use  all the energy of $N$
states  in the single coherent measurement.    The particles should be
divided into several groups  and the measurement should be sequentially
repeated several times. The accumulation  of information is  expressed
mathematically by the multiplication of corresponding distribution
functions (\ref{Leg}). This tends  to narrowing of the  posterior
probability distribution. The optimal regime for the $\theta$ detection
has been roughly estimated in Ref.\cite{acta}. Optimal number of
repetitions of the coherent measurement was expected to be
approximately
$n\approx 4$.

Let us interpret this interferometric measurement  in terms of the
spin estimation.  The significant difference between
spin 1/2 particles and photons   is connected with their
fermionic and bosonic nature.
The input wave function constructed  from fermions must be
``artificially'' symmetrized  with respect to all the distinguishable
particles. The input state $|N, 0\rangle $ is a
superposition involving all the combinations  of  $N/2$ particles with the
spin up and $N/2$ particles with the spin down. This $N$ particle state is
entangled.

\begin{figure}
     \epsfxsize=0.8\hsize \epsffile{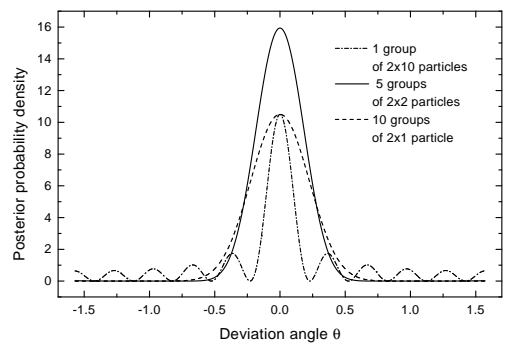} \caption{The
   posterior probability distribution provided that the exactly
   same number of spins ``up'' and ``down'' has been detected. The
   total number of particles is 20.} \label{fig2}
\end{figure}

\begin{figure}
     \epsfxsize=0.8\hsize \epsffile{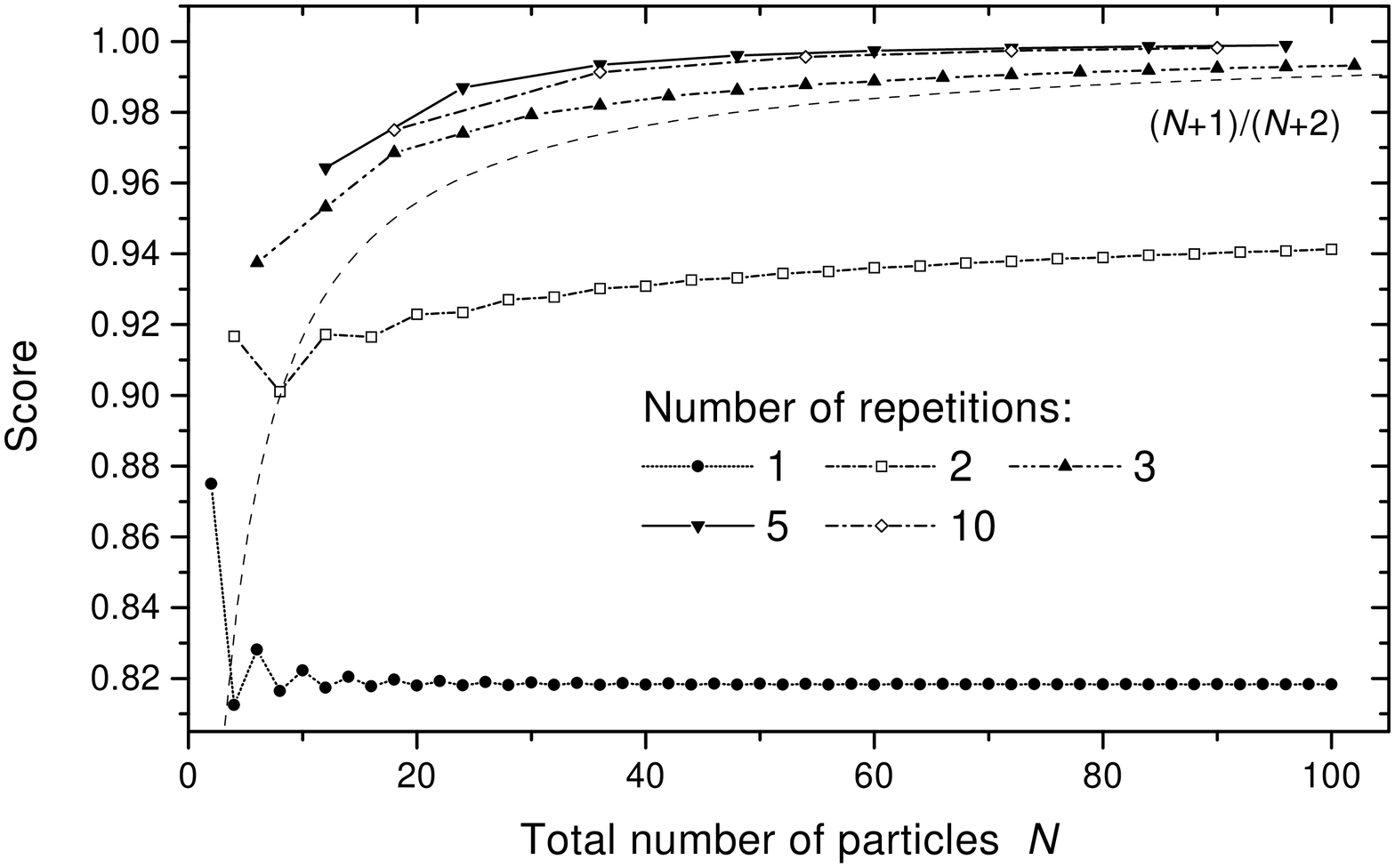} \caption{Ultimate
   values of score for several repetitions of coherent measurements
   on groups of particles.} \label{fig3}
\end{figure}

Let us analyse in detail the  the score for
such states. As before the most  favourable but still physically
feasible situation will be  interpreted as an ultimate bound of
the state detection.
Due to the symmetry of the problem, the best resolution may be
inferred provided that the state $|N,0\rangle $  feeds the SG
apparatus.
The results of such measurement will be deterministic in the case
when  the apparatus is  oriented along the spin state of each
particle. In this case a half of particles will be detected
with the spin ``up" and a half of particles with the spin
``down".  Assume now that this really happened.
%%%%%%%%
% We will not ask: ``How to do this ?" We will
% simply consider this ultimate situation
% because it obviously guarantees the
% best resolution.
%%%%%%%%
What can be
inferred from this event? Obviously, this might occure also in
the case  when the setting of SG differs  from the right spin
orientation by an angle  $\theta.$
The probability that  this happens  is proportional to the scattering
amplitude  (\ref{Leg}). The probability is sketched in Fig.~\ref{fig2}
-- as can be seen it shows oscillations.   The domain
is  restricted to the values $(0,\pi/2)$ since the method just
finds the axis but not its vector orientation.
The score is plotted in Fig.~\ref{fig3}. As shown it does not improve
with increasing number of particles. It is caused by the  heavy
tails  of the posterior distribution (Fig.~\ref{fig2}), which does not
contain the dominant  amount of the probability in the central
peak.  Nevertheless, there is a way how to suppress this undesired
behaviour. Provided that  the measurement  is repeated, the
corresponding  posterior distribution will be given by the
product of partial results. Again, the most favourable
situation  for spin estimation   is characterized by the
repetition of the same ``optimal''  result. This is again feasible, provided
that the true spin orientation and projection do not differ
significantly.
The resulting score in dependence on the total number of particles
is sketched in Fig.~\ref{fig3}. Notice, that SG measurement must be
done with $N/2$  particle state in this case, since the measurement
is repeated twice. This procedure may be further generalized
including an  arbitrary number of repetitions. As shown
in Fig~\ref{fig3}, the score  increases up to the 5 repetitions
and then starts to decrease.
This seems to be in good agreement
with the rough asymptotical analysis  presented in Ref.~\cite{acta},
where the optimal repetition rate has  been found as~4.

This result provides an ultimate bound in the following sense:
If the given input state
is measured  with the help of SG projections, the score
cannot be better than this  ultimate value. The question whether
this value may be attained is not answered here. Intuitively, in the
case of large number of particles it might be well approximated
by an adaptive scheme as in previous case of standard resolution.
However, the adaptive  scheme cannot be
applied to the first measurement. The results
may therefore appear as overestimated  in this case.
This explains why the analysis
applied here provides the score $S\approx 0.875$ for the state
$|\uparrow, \downarrow \rangle,$  whereas the result of Gisin and
Popescu \cite{gisin}
gives the value $0.789.$
In the asymptotical limit, the dispersion is approximately  given
by the relation
\begin{equation}
     D \approx \frac{\sqrt{4 n}}{N} = \frac{2\sqrt{5}}{N}.
\end{equation}
The optimal score reads asymptotically
\begin{equation}
  S \approx 1 - \frac{5}{N^2} .
\end{equation}

One may ask whether the required states containing half particles with
spins ``up'' and half with spins ``down'' (in an arbitrary direction)
can be prepared from a set of particles with all spins ``up'',
i.e. if it is possible to ``turn'' arbitrary quantum state into a state
orthogonal to it. The general answer is no -- the linearity of quantum
mechanics does not allow this.
Particularly in case of 1/2-spin particles, if one is able to do:
$| \uparrow \rangle \to | \downarrow \rangle$ and
$| \downarrow \rangle \to e^{i\varphi}| \uparrow \rangle$ then
for an arbitrary state it follows from linearity of quantum evolution that
$$ | \Psi \rangle =
 (\alpha | \uparrow \rangle  + \beta | \downarrow \rangle) \to
 (\alpha | \downarrow \rangle  + e^{i \phi} \beta | \uparrow \rangle).
$$
The last state is orthogonal to $ | \Psi \rangle$ if and only if
$\arg(\alpha) + \pi = \arg(\beta) +\phi$.
Thus in special situations (corresponding to ``real'' subspaces)
the mentioned transformation is possible
(e.g.\ when spin projections lay in a given plane, also for
linear polarizations of photon or for interferometry with fixed splitting
ratios and varied a phase difference only). But an arbitrary spin cannot
be turned to the perfectly opposite one.

\section{Conclusions}

The profound  analogy between interferometry and   spin estimation has been
addressed.
The performance  of both the schemes discussed has been
conditioned by the  realistic measurements only. As demonstrated, the
recently reported optimal   spin estimation corresponds to
the standard quantum limit characterized by the resolution
$1/\sqrt N.$  It may be achieved by coherent measurements and
well approximated by sequential ones. Beyond this regime, the
quantum theory admits the  resolution up to $1/N.$ Then,
however, quantum interferences must be employed. In the case of
a thought experiment with an ensemble of spin 1/2 particles it
requires  an entangled input state of $N$ particles. Besides
this,  optimal SG  detection must combine
advantages of both the coherent and sequential
measurements.
This example illustrates  complexity of the  optimal treatment
in estimation problems.

\section*{Acknowledgments}

Discussions with V. Bu\v{z}ek and D. Terno are gratefully acknowledged.
This work has been supported by the grant
VS 96028  and by the research project ``Wave and particle optics" of
the   Czech Ministry of Education
 and by the grant No. 19982003012 of  the Czech National Security Authority.

\end{document}